\documentclass[a4paper,twocolumn,11pt,accepted=2022-09-26]{quantumarticle}
\pdfoutput=1
\usepackage{graphicx, caption}
\usepackage{amsfonts}
\usepackage{amsmath,bm}
\usepackage{cite}
\usepackage{IEEEtrantools}
\usepackage{amssymb}
\usepackage{amsthm}
\usepackage{color}
\usepackage{mathtools}
\usepackage{braket}
\usepackage{enumitem}
\usepackage{multirow}
\usepackage{hyperref}
\usepackage{balance}

\newcommand{\CC}{\mathbb{C}}
\newcommand{\RR}{\mathbb{R}}
\newcommand{\tendsto}{\rightarrow}

\theoremstyle{plain}
\newtheorem{thm}{Theorem}

\newtheorem{defn}[thm]{Definition} 

\newtheorem{proposition}[thm]{Proposition}

\newtheorem{corollary}[thm]{Corollary}

\newtheorem*{remark}{Remark}

\newcommand\blfootnote[1]{%
  \begingroup
  \renewcommand\thefootnote{}\footnote{#1}%
  \addtocounter{footnote}{-1}%
  \endgroup
}

\title{Quantum Monte Carlo Integration: The Full Advantage in Minimal Circuit Depth}
\date{ }
\author[1,2]{Steven Herbert} 
\affil[1]{\textit{Quantinuum (Cambridge Quantum), Terrington House, 13-15 Hills Rd, Cambridge, CB2 1NL, UK}}
\affil[2]{\textit{Department of Computer Science and Technology, University of Cambridge, UK}}

\begin{document}

\maketitle
    \sloppy
\begin{abstract}
\noindent This\blfootnote{\noindent $^\dagger$ 
    Contact: Steven.Herbert@Quantinuum.com \\ After posting a pre-print of this paper, we were made aware of an existing work, Ref. \cite{blank2020quantumenhanced}, which proposes a somewhat similar decomposition of the Monte Carlo integral — although we note that Ref. \cite{blank2020quantumenhanced} lacks the detailed algorithmic specification and rigorous proof of performance that are central to the present work.} paper proposes a method of quantum Monte Carlo integration that retains the full quadratic quantum advantage, without requiring any arithmetic or quantum phase estimation to be performed on the quantum computer. No previous proposal for quantum Monte Carlo integration has achieved all of these at once. The heart of the proposed method is a Fourier series decomposition of the sum that approximates the expectation in Monte Carlo integration, with each component then estimated individually using quantum amplitude estimation. The main result is presented as theoretical statement of asymptotic advantage, and numerical results are also included to illustrate the practical benefits of the proposed method. The method presented in this paper is the subject of a patent application [Quantum Computing System and Method: Patent application GB2102902.0 and SE2130060-3].
\end{abstract}

 
 

\section{Introduction}
\label{intro}

\noindent Monte Carlo integration (MCI) solves the problem of estimating the mean of some probability distribution by averaging samples therefrom. This could either be an average of the samples themselves, or of the samples with some function applied first. It is known that a quadratic quantum speed-up in MCI is available \cite{MontanaroMC} by calling quantum amplitude estimation (QAE) \cite{brassard2000quantum} as a sub-routine. Such is the ubiquity of Monte Carlo methods throughout the physical, biological, data and information sciences, that this has, in turn, spawned a great deal of interest in \textit{quantum Monte Carlo integration} (QMCI) -- most notably in applications related to finance such as \textit{option pricing} \cite{an2020quantumaccelerated, QCfinance, egger2019credit, chakrabarti2020threshold, rebentrost2018quantum, kaneko2020quantum, Woerner2019, Rebentrost2018, financeppr, Miyamoto_2020, Kubo}.



However, the theoretical possibility of using quantum computing to perform MCI more efficiently has not masked a number of clear obstacles that are in the way of actually doing so in practice. Most notably the quantum advantage in MCI is an advantage in \textit{query} complexity: the number of uses of a quantum state that encodes the probability distribution in question will be fewer than the number of classical samples from that distribution -- but this immediately raises the question of how to prepare such a quantum state. Recently it was shown that there is no advantage when the oft-cited Grover-Rudolph method \cite{grover2002creating} is used to prepare states encoding probability distribution when a full audit of the \textit{computational} complexity is undertaken \cite{herbert2021problem}. 

Even should this problem be resolved, there are a number of reasons to be skeptical about the possibility of QMCI yielding a useful quantum advantage in the NISQ era. Most notably, QAE, as originally described, requires use of the quantum phase estimation (QPE), which is unlikely to be practical on near-term quantum hardware. On this front, though, there are reasons for optimism --  it has been shown that QPE can be omitted in favour of classical post-processing to extract an estimate of the statistical quantity of interest from measurements of the quantum state \cite{Suzuki_2020, grinko2019iterative, Aaronson_2020, nakaji2020faster}. Moreover, recent efforts have been made to further cut the required circuit depth by interpolating between classical and quantum MCI \cite{QCWpatent, giurgicatiron2020low}.

\begin{table*}[!t]
\begin{center}
  \begin{tabular}{l   l  l l }
     \textbf{Method} &   \textbf{Computes} &  \textbf{MSE} &  \textbf{Arithmetic} \\ \hline \hline
    Classical Monte Carlo Integration  &  $\mathbb{E}(f(X) g(Y))$ &  $\Theta(q^{-1})$ &  Classical  \\ 
    Quantum Monte Carlo Integration  &  $\mathbb{E}(f(X))$ &  $\Theta(q^{-2})$ &  Quantum \& classical \\ 
    Rescaled QMCI \cite{Stamatopoulos_2020, Woerner2019}  &  $\mathbb{E}(X)$ &  $\Theta(q^{-4/3})$ &  Classical only  \\ 
     Fourier QMCI (this paper)  & $\mathbb{E}(f(X) g(Y))$  &  $\Theta(q^{-2})$ &  Classical only \\ 
  \end{tabular}
\end{center}
\captionsetup{width=0.95\linewidth}
\caption{Comparison of different MCI estimation methods: $X$ and $Y$ are the values of different variables sampled from some multivariate probability distribution; $q$ is the number of samples (classical) or uses of a circuit that prepares a state encoding the probability distribution (quantum); and $f$ and $g$ are functions. We can see that our proposed method is the ``best of all worlds'': it is capable of computing the product of functions of two random variables, like classical MCI; and it achieves the full quantum speed-up, without requiring arithmetic operations to be performed quantumly -- something that none of the existing QMCI algorithms achieve. It is worth noting that there is some subtlety in the permissible function $f(.)$: in the case of conventional QMCI, $f(.)$ must be such that $0 \leq f(x) \leq 1$ for all $x$ -- with $x$ transformed accordingly if this is not already the case -- this restriction does not apply to $f(.)$ and $g(.)$ in Fourier QMCI; on the other hand, Fourier QMCI requires $f(.)$ to be in a form that can be decomposed into a Fourier series, which is not the case for conventional QMCI, and indeed the latter can even handle a quantum encoding of $f(.)$ (say from a prior quantum computation), as it is executed coherently, which is not the case for Fourier QMCI.}
\label{tab1}
\end{table*}

To a certain extent, however, these advances have served only to reveal further issues with NISQ compatible QMCI: specifically that the ``natural'' quantity to estimate on a quantum computer is:
\begin{equation}
\label{eq00}
\mathbb{E} \left( \sin^2 (X) \right) = \sum_{x } p(x) \sin^2 (x)
\end{equation}
where $X$ are samples from $p(x)$. Any other quantity, including the mean of $p(x)$ itself, will incur a significant amount of arithmetic to be performed on the quantum computer. Indeed, this cost is not unlike that of using QPE -- not a problem asymptotically, but almost certainly prohibitive in the NISQ era. The foremost proposal to work around the need for quantum arithmetic is showcased in \textit{Option Pricing using Quantum Computers} \cite{Stamatopoulos_2020} in which a technique proposed by Woerner and Egger \cite{Woerner2019} is used. Specifically, the technique shifts and rescales the support of the probability distribution such that for all $x \, : \, p(x) \neq 0$, $\sin^2(x + \pi/4) \approx x + 1/2$, thus enabling an estimate of the mean to be extracted. However, this method has an adverse effect on the rate of convergence, which can be seen as a false economy: the requirement of quantum arithmetic has been removed to reduce the circuit depth, but this comes at a cost of reduced convergence rate, and so correspondingly deeper circuits will be required to achieve the specified estimation accuracy.\\
\indent In this article we propose an alternative solution that still eliminates the need for quantum arithmetic, but also upholds the full quadratic quantum advantage. Instead of \textit{squeezing} the support of the probability distribution to correspond to an approximately linear region of a trigonometric function, instead we propose that the function applied to the samples is \textit{extended} as a periodic, piecewise function. This periodic function can be decomposed as a Fourier series, whose various components have means that can be estimated individually and then recombined. With judicious choices about how to do this Fourier series decomposition, we show that the quadratic quantum advantage is indeed retained.\\
\indent Furthermore, by performing the QMCI in this way, there is no ``favoured'' quantity to compute (in the sense of (\ref{eq00})), and computation of the mean of any function applied to the samples is equally easy in principle. This immediately suggests a way in which an actual computational advantage, as opposed a theoretical advantage in query complexity, may be realised using QMCI. Quantum advantage can be realised using the proposed method if: (a) there exists some probability distribution that can be exactly prepared as a quantum state in shallow-depth; (b) a function is then applied to the random samples such that the computed mean is a quantity that cannot be analytically evaluated. This is a clear and useful distinction over, for example, the method proposed by Stamatopoulos \textit{et al} where the distribution of interest must be directly encoded into the quantum state  \cite{Stamatopoulos_2020}. \\
\indent Finally, it is worth highlighting that our proposed method of QMCI can also be used to compute the expectation of the product of functions of two correlated random variables, again with the full quadratic quantum advantage and without requiring quantum arithmetic. This may find application in, for example, the computation of correlations and covariances of random variables from multivariate probability distributions. \\
\indent Table~\ref{tab1} summarises the benefits of the method of QMCI proposed here compared to existing alternatives. Convergence is measured in mean-squared error (MSE) as a function of the number of samples from the probability distribution, $q$. The table includes classical MCI for comparison, but it should be noted that there are classical alternatives, for example quadrature and quasi-Monte Carlo methods. Ostensibly, these appear to have favourable performance (achieving MSE convergence proportional to $q^{-2}$), however they suffer the curse of dimensionality: the complexity grows exponentially with the number of dimensions. Hence it is noteworthy that our proposed method applies to random variables marginalised out from multivariate probability distributions as, for a sufficiently large number of dimensions, classical MCI will be the most appropriate classical algorithm to benchmark against.\\
\indent The remainder of this paper is organised as follows: in Section~\ref{prelim} the problem at hand is specified, the relevant definitions given, and precise details of the motivation presented; in Section~\ref{main} the main theoretical result is given; in Section~\ref{corr} the result pertaining to the expectation of products of random variables is given as a Corollary, whose proof is then given in Appendix~\ref{appA}; in Section~\ref{num} a simulated example is given as a numerical proof of principle; and in Section~\ref{disc} the results of the paper are discussed. The method presented in this paper is the subject of a patent application \cite{sjhpatent}.

\section{Preliminaries and Motivation}
\label{prelim}

\noindent Suppose we have the $d$-dimensional discrete probability distribution $p$ (that is a probability distribution over some finite subset of $\mathbb{R}^d$). The points of non-zero probability mass are spaced at constant intervals and for simplicity, we let the number of points along each axis be a power of 2, say $2^{N_j}$ for the $j^{th}$ dimension (we can ``zero-pad'' so there is no loss of generality in this assumption). The probability distribution is encoded in a quantum state, $\ket{p}$, of the form:
\begin{equation}
\label{eqn10}
\ket{p} = \sum_{\substack{x^{(1)} \in \{ 0,1 \}^{N_1} \\ \vdots \\ x^{(d)} \in \{ 0,1 \}^{N_d} }}\sqrt{p(x^{(1)} , \dots , x^{(d)})} \ket{x^{(1)}  \dots x^{(d)}}  
\end{equation}
and we assume that the quantum algorithm has access to a circuit, $P$, that prepares $\ket{p}$, ie, $\ket{p} = P\ket{\mathbf{0}}$ (where $\ket{\mathbf{0}}$ denotes the tensor product of an appropriate number of qubits in the $\ket{0}$ state).

QMCI estimates the expectation of some bounded function, $f(x)$, of samples from the marginal distribution of a single dimension of the multivariate probability distribution. This dimension must therefore be specified -- let this be the $i^{th}$ dimension -- so that QMCI returns an estimate of the mean of the probability distribution implicitly defined by $f(x)$ applied to samples, $X$, from $p(x^{(i)})$, that is an estimate of:
\begin{equation}
\mathbb{E}(f(X)) =  \sum_{x^{(i)}} f(x^{(i)}) p(x^{(i)}) 
\end{equation}
To achieve this, the support of $p(x^{(i)})$ is also taken as an input to the QMCI algorithm: this is given as the value that the first point of the distribution corresponds to, $x_l$ and the spacing interval, $\Delta$. It is also useful to express the maximum value of the support of the distribution, $x_u = x_l + (2^{N_i} - 1) \Delta$.

In order to obtain a quantum advantage using the QMCI method in this paper, we further require that $f(x)$ is continuous, has continuous first derivative and bounded piecewise-continuous second and third derivatives over the support of $p(x^{(i)})$ (`the support of $p(x^{(i)})$' meaning $x_l \leq x^{(i)} \leq x_u$). QMCI can be achieved by constructing a matrix $A = R (P \otimes I)$ where 
\begin{align}
R & \ket{x^{(1)}  x^{(2)} \dots x^{(d)}}   \! \ket{0}   \nonumber  \\ 
= & \ket{x^{(1)} x^{(2)} \! \dots \! x^{(d)}} \! \left(\! \sqrt{1-f(x^{(i)})} \! \ket{0} \!+ \! \sqrt{f(x^{(i)})} \!\ket{1}\! \right) \nonumber
\end{align}
Thus $A \ket{\mathbf{0}} \! = \! \cos \theta_a \ket{\Psi_0} \!\ket{0} + \sin \theta_a \ket{\Psi_1} \! \ket{1} $ where $\ket{\Psi_0}$ and $\ket{\Psi_1}$ are just some quantum states and $\sin^2 \theta_a = \sum_{x^{(i)}} f(x^{(i)}) p(x^{(i)})$, ie, the expectation we are seeking (see Ref. \cite{Suzuki_2020} for a more detailed exposition of this), which can be estimated by calling some QAE algorithm as a subroutine\footnote{Strictly speaking, it is required that $0 \leq f(x) \leq 1$ for all $x$, and so if this is not the case then some transformation must be performed. For instance, in the case where $f(x)=x$ (that is the mean is being estimated) then a simple affine transformation suffices, but more generally this may be trickier.}.
\begin{defn}
\label{def1}
A QAE algorithm takes as an input a circuit $A$ where $A \ket{\mathbf{0}} \! = \! \cos \theta_a \ket{\Psi_0} \! \ket{0} + \sin \theta_a \! \ket{\Psi_1} \ket{1}$, and returns an estimate of $ s = \sin^2 \theta_a$. The allowed number of uses, denoted ``$q$'', of the circuit $A$ is given as an input to the QAE algorithm. 
\end{defn}
\noindent From this definition, we can write any QAE algorithm as a function in pseudocode form:
\begin{equation}
\hat{s} = \texttt{QAE}(A, q)
\end{equation}
where $\hat{s}$ expresses the fact that this is an estimate of $s$. It is also useful to formally define the MSE, $\hat{\epsilon}^2$:
\begin{equation}
\hat{\epsilon}^2 = \mathbb{E}((\hat{s} - s)^2)
\end{equation}
The convergence of any QAE algorithm can be expressed as:
\begin{equation}
\label{qae-converge}
\hat{\epsilon}^2(q) \in \Theta(q^{-\lambda})
\end{equation}
for some $\lambda$, and the method proposed in this paper assumes that the convergence rate of the QAE algorithm as a function of the number of uses of $A$ is known. That is, there exists a constant $k_1$ and we know the constant $\lambda$ such that:
\begin{equation}
\label{eq50}
\hat{\epsilon}^2(q) \leq k_1 q^{-\lambda}
\end{equation}
A classical MCI algorithm can be used to perform QAE, in the sense of the above definition, and in this case $\lambda=1$, whereas typical QAE algorithms have $\lambda=2$. However, the method we propose here works for any QAE algorithm, including a proposed method that interpolates between the classical and quantum (so $1 \leq \lambda \leq 2$) \cite{bouland2020prospects, QCWpatent, giurgicatiron2020low}.\\
\indent QAE algorithms use the circuit $A$ (and hence $P$) to build a ``Grover iterate'' circuit, 
\begin{equation}
Q = -A S_0 A^{\dagger} S_{\chi}
\end{equation}
where $S_0$ and $S_{\chi}$ do not depend on $P$ or $A$. Further details do not directly concern us here -- apart from that circuit depth of QMCI is usually counted in number of sequential uses of $Q$  \cite{brassard2000quantum, Suzuki_2020}.\\

\subsection*{\hspace{-0.55cm} Problems with existing methods}

We can see that performing QAE when $A = R (P \otimes I)$ returns an estimate of $s = \sum f(x) p(x)$, and thus \textit{does} yield a quantum speed-up. However, this comes at the cost of using the circuit $R$, which encodes the function applied to the random samples, and in general this will be prohibitively complex -- even when QPE-free forms of QAE are used. Even the (classically) trivial case where $f(x) = x$ (ie, we are just finding the mean of $p(x)$) would entail significant amounts of arithmetic to be performed quantumly. An important exception is the special case where $f(x) = \sin^2 (mx + c)$ for some constants $m$ and $c$, which can be achieved by a bank of $R_y$ rotation gates, as shown in Fig.~\ref{fig0}. 

\section{QMCI of a Single Random Variable}
\label{main}

\begin{figure}[!t]
\centering
\includegraphics[width = 0.95\linewidth]{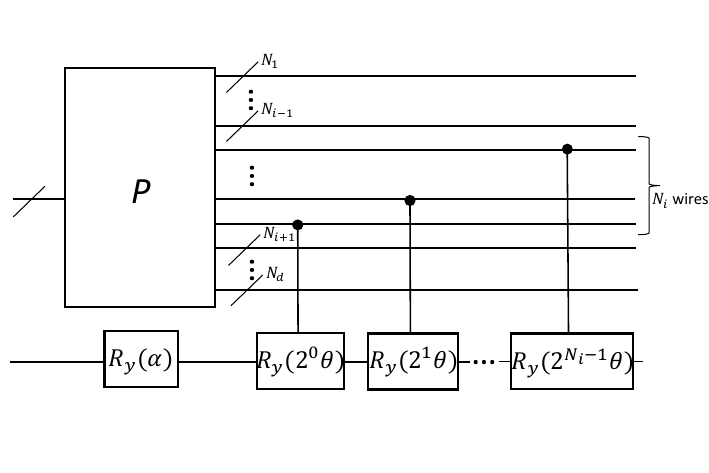} %
\captionsetup{width=0.95\linewidth}
\caption{Circuit diagram of $A(P, i, \beta, n, \omega)$, where $\alpha = n \omega x_l - \beta$ and $\theta = n \omega \Delta $, in which $x_l$ and $\Delta$ are the first point of probability mass and spacing (as defined in Section~\ref{prelim}) for the $i^{th}$ dimension of $P$.}
\label{fig0}
\end{figure}

\noindent The main result of the paper requires QAE to be performed on the quantum circuit $A(P, i, \beta, n, \omega)$ given in Fig.~\ref{fig0}. $P$ is a circuit that prepares from $\ket{\mathbf{0}}$ a quantum state encoding a multivariate probability distribution, as in (\ref{eqn10}), and $i$ denotes the dimension of interest (we assume that the support of the $i^{th}$ dimension is known, in terms of the value of the first point of non-zero probability mass, $x_l$ and spacing between points of probability mass, $\Delta$); $n$  is an integer and $\omega$ and $\beta$ real numbers. 
\begin{proposition}
\label{lem1}
Let $\mathtt{QAE}(A, q)$ be a QAE algorithm in the sense of Definition~\ref{def1}, which has MSE convergence parameterised by $\lambda$ as in (\ref{qae-converge}).
\begin{enumerate}[label=\roman*., leftmargin=*]
\item $1 - 2\mathtt{QAE}(A(P, i, 0, n, \omega),q)$ is an estimate of:
\begin{equation}
\sum_{x^{(i)}} p(x^{(i)})  \cos (n \omega x^{(i)} )
\end{equation}
with MSE $\in \Theta(q^{-\lambda})$.\\
\item $1 - 2\mathtt{QAE}(A(P, i, \pi/2, n, \omega),q)$ is an estimate of:
\begin{equation}
\sum_{x^{(i)}} p(x^{(i)})  \sin (n \omega x^{(i)}) 
\end{equation}
with MSE $\in \Theta(q^{-\lambda})$.
\end{enumerate}
\end{proposition}

\begin{proof}
The function of circuit $A$ is to apply an $R_y$ rotation to the last qubit that is proportional to some constant plus the value of the $i^{th}$ register of $P$. Recall that the circuit $P$, when applied to $\ket{\mathbf{0}}$, encodes a multivariate probability distribution in the computational basis states of a quantum state, as defined in (\ref{eqn10}), thus from Fig.~\ref{fig0} we can see that, when $\beta=0$, $A\ket{\mathbf{0}}$ prepares the state:
\begin{align}
 A\ket{\mathbf{0}} = & \sum_{x^{(1)} \dots x^{(d)} }  \sqrt{p(x^{(1)},  \dots , x^{(d)})}  \ket{x^{(1)}  \dots x^{(d)}}   \nonumber \\
&\,\,\,\,\,\,\,\,\left( \! \cos \left( \frac{n \omega x^{(i)} }{2} \right) \ket{0} \! + \! \sin \left( \frac{n \omega x^{(i)} }{2} \right) \ket{1} \! \right)
\end{align}
and thus the amplitude of the $\ket{1}$ component of the final qubit is:
\begin{align}
 \sum_{x^{(1)} \dots x^{(d)} } &  p(x^{(1)},   \dots , x^{(d)})   \sin^2 ( n \omega x^{(i)}/2 ) \nonumber  \\
    & = \sum_{x^{(i)}} p(x^{(i)})   \sin^2 ( n \omega x^{(i)}/2 ) \nonumber  \\
    & = \sum_{x^{(i)}} \frac{1}{2}p(x^{(i)}) (1-  \cos (n \omega x^{(i)})) \nonumber \\
    & = \frac{1}{2}\left(1- \sum_{x^{(i)}} p(x^{(i)})  \cos (n \omega x^{(i)} )\right)
\end{align}
where $p(x^{(i)})$ is the marginal distribution -- note that the marginalisation has been achieved automatically.

Thus $\texttt{QAE}(A(P, i, 0,n,\omega),q)$ estimates the value of $0.5(1-  \sum_{x^{(i)}} p(x^{(i)})  \cos (n \omega x^{(i)} ))$ with MSE $\in \Theta(q^{-\lambda})$, by (\ref{qae-converge}). So it follows that $1 - 2\texttt{QAE}(A(P,i,0,n,\omega),q)$ therefore estimates $ \sum_{x^{(i)}} p(x^{(i)})  \cos (n \omega x^{(i)} )$, and as this differs from the original estimate only by being shifted and scaled by a constant factor, the estimation accuracy remains such that MSE $\in \Theta(q^{-\lambda})$.

Turning now to the second part of the Proposition, by the same reasoning as before, when $\beta = \pi / 2$, $A\ket{\mathbf{0}}$ prepares the state:
\begin{align}
A\ket{\mathbf{0}} & \! = \!   \sum_{x^{(1)} \dots x^{(d)} } \sqrt{p(x^{(1)},  \dots , x^{(d)})} \ket{x^{(1)}  \dots x^{(d)}}   \nonumber \\
& \left( \! \cos \! \left( \! \frac{n \omega x^{(i)} }{2} \! - \! \frac{\pi}{4}\! \right)\! \ket{0} \!+ \! \sin \! \left(\!  \frac{n \omega x^{(i)} }{2} \! - \!\frac{\pi}{4} \right)\! \ket{1} \! \right)
\end{align}
and thus the amplitude of the $\ket{1}$ component of the final qubit is:
\begin{align}
    \sum_{x^{(i)}} & p(x^{(i)})   \sin^2 \left( \frac{n \omega x^{(i)} }{2} - \frac{ \pi }{4} \right) \nonumber \\
    & = \sum_{x^{(i)}} \frac{1}{2}p(x^{(i)}) \left( 1-  \cos \left(  n \omega x^{(i)}  - \frac{ \pi }{2} \right) \right) \nonumber \\
    & = \frac{1}{2}\left(1- \sum_{x^{(i)}} p(x^{(i)})  \sin (n \omega x^{(i)}) \right)
\end{align}
Thus $1 - 2\texttt{QAE}(A(P, i, \pi/2, n , \omega), q)$ gives an estimate of $\sum_{x^{(i)}}  p(x^{(i)})  \sin (n \omega x^{(i)} )$ with MSE $\in \Theta( q^{-\lambda})$.
\end{proof}

\noindent Proposition~\ref{lem1} now allows us to proceed to the main result of the paper.

\begin{thm}
The quantity, $\mu = \mathbb{E}(f(X))$, where $X \sim p(x^{(i)})$ and $f$ is a function that is continuous in value and first derivative and whose second and third derivatives are piecewise-continuous and bounded, can be estimated with MSE $\in \Theta(q^{-\lambda})$, where $q$ is the number of uses of a circuit preparing the quantum state $\ket{p}$ (ie, as defined in (\ref{eqn10})) and $\lambda$ is the convergence rate of some QAE subroutine (ie, as defined in (\ref{eq50})) which operates on circuits of the form defined in Fig.~\ref{fig0}.
\end{thm}
\begin{remark}
This result substantiates the ``best of all worlds'' claim made in Section~\ref{intro}: the convergence of the QMCI estimate is equal to that of the underlying QAE algorithm; whilst the circuit in Fig.~\ref{fig0} requires only a single bank of rotation gates to construct $A$ from $P$, which is minimal. That is, if the probability distribution of interest is over $2^{N_i}$ points then at least $N_i$ two-qubit gates are required in order that the amplitude of a further qubit is equal to the mean of some function the probability distribution (ignoring unnatural cases where some of the data-points of the probability distribution are omitted altogether).
\end{remark}

\begin{proof}
We prove the theorem by giving an explicit construction. The proof strategy is to first build a periodic function, $\mathrm{f}(x)$, from $f(x)$; then to show that the expectation can be approximated by performing QAE on the components of a Fourier series decomposition of $\mathrm{f}(x)$ using Proposition~\ref{lem1}; and finally to show that the total number of uses of $P$ can be distributed amongst the Fourier series components such that the claimed overall convergence rate holds.\\
\indent First, we build a periodic piecewise function $\mathrm{f}(x)$ such that $\mathrm{f}(x) = f(x)$ over the support of $p(x)$; $\mathrm{f}(x)$ is also constructed such that it is continuous in value and first derivative, and has second and third derivatives that are piecewise-continuous and bounded. For example, some $\mathrm{f}(x)$ of the following form, which repeats with period $x_{\tilde{u}} - x_l$, is suitable in general:
\begin{equation}
\mathrm{f}(x) = \begin{cases} f(x)  &\mbox{if } x_l \leq x < x_u  \\
\tilde{f}(x) & \mbox{if } x_u \leq x < x_{\tilde{u}}  
\end{cases}
\end{equation}
where $x_{\tilde{u}} > x_u$ and $\tilde{f}(x)$ is itself sufficiently smooth (that is, continuous in value and first derivative, and has second and third derivatives that are piecewise-continuous and bounded) and chosen such $f(x_l) = \tilde{f}(x_{\tilde{u}}  )$,  $f'(x_l) = \tilde{f}'(x_{\tilde{u}}  )$, $f(x_u) = \tilde{f}(x_u  )$ and $f'(x_u) = \tilde{f}'(x_u  )$ (where $f'$, $\tilde{f}'$ denotes differentiation w.r.t. $x$). It is easy to see that a cubic will always suffice to fit such a $\tilde{f}(x)$, and on occasion a simpler function may do. As $\mathrm{f}(x)$ is periodic, it has a Fourier series:
\begin{equation}
\mathrm{f}(x) = c + \sum_{n=1}^{\infty} a_n \cos (n \omega x) + b_n \sin (n \omega x)
\end{equation}
where $\omega = 2 \pi / T$ and $T$ is the period of the periodic piecewise function. By construction, $\mathrm{f}(x)$ is continuous in value and first derivative and has second and third derivatives that are piecewise-continuous and bounded, and so the Fourier series coefficients decay as $1/n^3$ or faster\footnote{This is a relatively commonly-used result, however there does not appear to be a standard citation -- so a theorem to this effect is given in Appendix~\ref{app-zen}} and thus we can write:
\begin{equation}
\mathrm{f}(x) = c + \sum_{n=1}^{\infty} \frac{1}{n^3} \left( \tilde{a}_n \cos (n \omega x) + \tilde{b}_n \sin (n \omega x) \right)
\end{equation}
where $|\tilde{a}_n|, \, |\tilde{b}_n| \in \mathcal{O}(1)$. As $\mathrm{f}(x) = f(x)$ whenever $p(x^{(i)}) \neq 0$, we can express:
\begin{align}
    \mu = & \sum_{x^{(i)}} p(x^{(i)}) f(x^{(i)}) \nonumber \\
      = & \sum_{x^{(i)}} p(x^{(i)}) \mathrm{f}(x^{(i)}) \nonumber \\
     = & \sum_{x^{(i)}} p(x^{(i)}) \Bigg(  \sum_{n=1}^\infty \frac{1}{n^3} \left( \tilde{a}_n \cos (n \omega x^{(i)}) \right.  \nonumber \\
     & \,\,\,\,\,\,\,\,\,\,\,\,\,\,\,\,\,\,\,\,\,\,\,\,\,\,\,\,\,\,\,\,\,\,\,\,\,\,\,\,  \left.+ \tilde{b}_n \sin (n \omega x^{(i)}) \right) + c \Bigg) \nonumber \\
     = & c + \sum_{n=1}^\infty \frac{\tilde{a}_n}{n^3} \, \left( \sum_{x^{(i)}}  p(x^{(i)})  \cos( n \omega x^{(i)}) \right) \nonumber \\
     \label{eqeq170}
     & \,\,\, + \sum_{n=1}^\infty \frac{\tilde{b}_n}{n^3} \, \left( \sum_{x^{(i)}}  p(x^{(i)})  \sin ( n \omega x^{(i)}) \right)
\end{align}
To estimate $\mu$, we can thus estimate each of the parenthesised sums in the final line of (\ref{eqeq170}) individually. These can be estimated using QAE according to the procedure described in Proposition~\ref{lem1}. To analyse the overall accuracy of the consequent estimate of $\mu$, we first define:
\begin{align}
\label{eq-lab1}
    \mu^{(a)}_n = & \sum_{x^{(i)}} p(x^{(i)})  \cos (n \omega x^{(i)}) \\
    \mu^{(b)}_n = & \sum_{x^{(i)}} p(x^{(i)})  \sin (n \omega x^{(i)}) 
\end{align}
and further define $\hat{\mu}^{(a)}_n$ and $\hat{\mu}^{(b)}_n $ as the estimates of $\mu^{(a)}_n$ and $\mu^{(b)}_n$ respectively such that:
\begin{align}
    \mu^{(a)}_n & = \hat{\mu}^{(a)}_n + \nu^{(a)}_n + \epsilon^{(a)}_n  \\
    \mu^{(b)}_n & = \hat{\mu}^{(b)}_n + \nu^{(b)}_n +  \epsilon^{(b)}_n  
\end{align}
where $ \nu^{(a)}_n$ and $ \nu^{(b)}_n$ are biases (noting that the possibility of biased estimation has not been precluded) and $\epsilon^{(a)}_n$ and $\epsilon^{(b)}_n$ are zero-mean random variables. 
The inaccuracy of estimating $\mu$ in this way does not solely arise because of the inaccuracy of the estimates of each term in (\ref{eqeq170}), we must also address the fact that the Fourier series is an infinite sum, and so obviously we cannot estimate all of the terms in practice. Therefore we truncate at some $n_{max}$. That is, our estimate, $\hat{\mu}$, of $\mu$ is:
\begin{equation}
    \hat{\mu} = c + \sum_{n=1}^{n_{max}} \frac{\tilde{a}_n}{n^3} \hat{\mu}^{(a)}_n + \frac{\tilde{b}_n}{n^3} \hat{\mu}^{(b)}_n
\end{equation}
and to keep track of the error incurred by this truncation, we define the value of the truncated part of the sum:
\begin{align}
\label{rlab10}
    \eta =  \sum_{n=n_{max}+1}^\infty & \frac{\tilde{a}_n}{n^3} \, \left( \sum_{x^{(i)}} \cos( n \omega x^{(i)}) p(x^{(i)}) \right) \nonumber \\
    & + \frac{\tilde{b}_n}{n^3} \, \left( \sum_{x^{(i)}} \sin( n \omega x^{(i)}) p(x^{(i)}) \right)
\end{align}
and it is also convenient to define:
\begin{align}
\label{eq-lab2}
    \nu^{(a)} & = \sum_{n=1}^{n_{max}}  \frac{\tilde{a}_n}{n^3} \nu^{(a)}_{n} \\
    \label{eq-lab3}
    \nu^{(b)} & = \sum_{n=1}^{n_{max}}   \frac{\tilde{b}_n}{n^3} \nu^{(b)}_{n}
\end{align}
Substituting (\ref{eq-lab1}) -- (\ref{rlab10}) into (\ref{eqeq170}) and also using the definitions (\ref{eq-lab2}) and (\ref{eq-lab3}) we can express the actual mean, $\mu$, in terms of our estimate of the mean, $\hat{\mu}$:
\begin{align}
    \mu & = \hat{\mu} \! + \! \sum_{n=1}^{n_{max}}  \frac{\tilde{a}_n}{n^3} \! \left( \nu^{(a)}_n +  \epsilon^{(a)}_n \right) + \frac{\tilde{b}_n}{n^3} \! \left( \nu^{(b)}_n + \epsilon^{(b)}_n \right) \! + \eta \nonumber \\
    & =  \hat{\mu} + \nu^{(a)} + \nu^{(b)} + \sum_{n=1}^{n_{max}}  \frac{\tilde{a}_n}{n^3}   \epsilon^{(a)}_n  + \frac{\tilde{b}_n}{n^3} \epsilon^{(b)}_n  + \eta
\end{align}
from which we can express the MSE:
\begin{align}
\hat{\epsilon}^2  = & \mathbb{E}((\mu - \hat{\mu})^2) \nonumber \\
 = &    \mathbb{E} \! \left( \! \left( \! \nu^{(a)} \!+\! \nu^{(b)}\! +\! \eta\!  +\! \sum_{n=1}^{n_{max}}  \frac{\tilde{a}_n}{n^3} \epsilon^{(a)}_n\! + \! \frac{\tilde{b}_n}{n^3} \epsilon^{(b)}_n\!  \right)^2 \right)  \nonumber \\
\label{eqn160}   
  = &    (\nu^{(a)} + \nu^{(b)})^2   + \eta^2  +  (\nu^{(a)} + \nu^{(b)}) \eta \nonumber \\
& \!+ \! \sum_{n=1}^{n_{max}} \! \left(\frac{\tilde{a}_n}{n^3} \right)^2 \! \mathbb{E} \! \left( ( \epsilon^{(a)}_n)^2 \right)  \! + \!\left(\frac{\tilde{b}_n}{n^3} \right)^2  \mathbb{E} \left( ( \epsilon^{(b)}_n)^2 \right) 
\end{align}
using the fact that $\epsilon^{(a)}_n$ and $\epsilon^{(b)}_n$ are zero-mean by design, and also that estimates corresponding to the various terms in the Fourier series are independent. 

To prove the theorem it is now incumbent upon us to show that the number of uses of $P$ assigned to the QAE estimation of each of the terms in (\ref{eqeq170}) and $n_{max}$ can be chosen such that the claimed convergence holds. To do this, we first let $q_n$ be the number of uses of $P$ allowed in the QAE estimate of \textit{each} of the sine and cosine components of the $n^{th}$ term in the Fourier series such that:
\begin{equation}
\label{eqn161}
q_n = q_0 n^{-\kappa}
\end{equation}
for some constants $q_0$ and $\kappa$ the latter of which we will later give numerical value. Recalling that $|\tilde{a}_n|, \, |\tilde{b}_n| \in \mathcal{O}(1)$, we can define a constant $k_2$ such that $k_2 \geq |\tilde{a}_n|, \, |\tilde{b}_n|$ (in the following analysis we will need to introduce further new constants, and will henceforth do so without explicit announcement by using $k$ with subscript incremented relative to the most recently introduced constant). We now address the terms in (\ref{eqn160}). First, we note that the magnitude of the bias, $\nu^{(a)}_n$, is upper-bounded by the square root of the MSE, which has been bounded in (\ref{eq50}), so (starting with (\ref{eq-lab2}) and also using (\ref{eqn161}), and $k_2 \geq |\tilde{a}_n|$ as defined above):
\begin{align}
    |\nu^{(a)}| & \leq k_2 \sqrt{k_1} \sum_{n=1}^{n_{max}} \frac{1}{n^3} q_n^{-\lambda/2}  \nonumber \\
    & \leq k_2 \sqrt{k_1} q_0^{-\lambda/2} \sum_{n=1}^{\infty} n^{\kappa \lambda/2 -3} \nonumber \\
    \label{eqn210}
    & = k_3 q_0^{-\lambda/2} 
\end{align}
so long as $\kappa$ is chosen such that $\kappa < 4/\lambda$ and the sum therefore converges. In exactly the same way we get:
\begin{equation}
\label{eqn210a}
|\nu^{(b)}|   \leq k_3 q_0^{-\lambda/2} 
\end{equation}
Turning now to the term (of  (\ref{eqn160})), $\mathbb{E} \left( ( \epsilon^{(a)}_n)^2 \right)$, which is the mean-squared error of the unbiased part of the estimate of $\mu^{(a)}_n$, and is therefore upper-bounded by $k_1 q_n^{-\lambda}$ (this would hold with equality for an unbiased estimator and be strictly less for an estimator with non-zero bias). So we get (again using (\ref{eq50})):
\begin{align}
\sum_{n=1}^{n_{max}} \left( \frac{\tilde{a}_n}{n^3} \right)^2  \mathbb{E} \left( ( \epsilon^{(a)}_n)^2 \right) & \leq k_1 \sum_{n=1}^{n_{max}} \left( \frac{\tilde{a}_n}{n^3} \right)^2 q_n^{-\lambda}  \nonumber \\
    & \leq  k_1 k_2^2 q_0^{-\lambda} \sum_{n=1}^{\infty} n^{\kappa \lambda-6}  \nonumber \\
    \label{eqn230}
    & = k_4 q_0^{-\lambda}
\end{align}
where the previously specified condition, $\kappa < 4/\lambda$, suffices to ensure that this sum also converges. Likewise we get
\begin{equation}
\label{eqn230a}
\sum_{n=1}^{n_{max}} \left( \frac{\tilde{b}_n}{n^3} \right)^2  \mathbb{E} \left( ( \epsilon^{(b)}_n)^2 \right) \leq k_4 q_0^{-\lambda}
\end{equation}
The final quantity in  (\ref{eqn160}) that we must address is the value of the truncated term, $\eta$, defined in (\ref{rlab10}). The magnitudes of $\cos(n \omega x^{(i)})$ and $\sin(n \omega x^{(i)})$ are bounded above by $1$ for all $x^{(i)}$, and so 
\begin{align}
    -1 \leq \sum_{x^{(i)}} \cos( n \omega x^{(i)}) p(x^{(i)}) & \leq 1 \\
    -1 \leq \sum_{x^{(i)}} \sin( n \omega x^{(i)}) p(x^{(i)}) & \leq 1
\end{align}
and thus from (\ref{rlab10}), using $k_2 \geq |\tilde{a}_n|, |\tilde{b}_n|$ (by definition) we have
\begin{align}
    | \eta |& \leq \sum_{n=n_{max}+1}^\infty \frac{| \tilde{a}_n | + | \tilde{b}_n | }{n^3} \nonumber \\
    & \leq 2k_2 \sum_{n=n_{max}+1}^\infty \frac{1}{n^3} \nonumber \\
    & \leq 2k_2 \int_{n_{max}}^\infty \frac{1}{x^3} \, \mathrm{d} x \nonumber \\
    \label{eqn2300}
    & = \frac{ k_2 }{n_{max}^2}
\end{align}
Putting this all together by substituting (\ref{eqn210}), (\ref{eqn210a}), (\ref{eqn230}), (\ref{eqn230a}) and (\ref{eqn2300}) into (\ref{eqn160}) we get:
\begin{equation}
\label{eqn2400}
\hat{\epsilon}^2 \leq 4 k_3^2 q_0^{- \lambda} + \frac{k_2^2}{n^4_{max}} + \frac{2k_2 k_3 q_0^{-\lambda / 2}}{n^2_{max}} + 2 k_4 q_0^{-\lambda} 
\end{equation}
From which we can see that in order to achieve $\hat{\epsilon}^2 \propto q_0^{-\lambda}$ it suffices to choose $n_{max}$ such that:
\begin{equation}
\label{eqn2500}
    n_{max} = \lceil q_0^{\lambda/4} \rceil
\end{equation}
We must now count up the total number of uses of $P$. Recall that we have assigned $q_n$ uses of $P$ to \textit{each} of the sine and cosine components of the $n^{th}$ term in the Fourier series, thus we have $\sum_{n=1}^{n_{max}} 2 q_n$ uses of $P$. However, in the above analysis we have allowed $q_n$ to take non-integer values, but in practice we must round to an integer. To ensure that the claim of convergence holds, we round up in each case. If all $n_{max}$ Fourier components require $q_n$ to be rounded up by an amount approaching $1$ for each of the sine and cosine components then we will require an extra $2n_{max}$ uses of $P$ in the worst case. Thus we can evaluate the total number of uses of $P$:
\begin{align}
q_{tot} & = \sum_{n=1}^{n_{max}} 2 q_n + 2 n_{max} \nonumber \\
& = 2 q_0 \sum_{n=1}^{n_{max}} n^{-\kappa} + 2 \lceil q_0^{\lambda/4} \rceil \nonumber \\
& \leq 2 q_0 \sum_{n=1}^{\infty} n^{-\kappa} + 2 \lceil q_0^{\lambda/4} \rceil
\end{align}
From (\ref{eqn210}) we require that $\kappa < 4/ \lambda$, and we also have that in practice $ 1 < \lambda \leq 2$ (ie, at $\lambda \leq 1$ there is no quantum advantage, and $\lambda = 2$ is the Heisenberg limit). Thus choosing any $\kappa < 2$ always suffices, and so we let $\kappa = 2 - \delta$ for some $0 < \delta < 1$. As $\lambda \leq 2$, we have that $\lambda / 4 \leq 1/2$, and thus we get:
\begin{align}
q_{tot} &\leq 2 q_0 \sum_{n=1}^{\infty} n^{-2+ \delta} + 2 q_0^{1/2} \nonumber \\
& \in \Theta(q_0)
\end{align}
because the sum converges when $0 < \delta < 1$. Thus we have shown that the total number of queries is proportional to $q_0$. Putting this together with $\hat{\epsilon}^2  \propto q_0^{- \lambda}$ from (\ref{eqn2400}) and (\ref{eqn2500}) we can thus see that $\hat{\epsilon}^2 \in \Theta(q_{tot}^{-\lambda})$, proving the theorem.
\end{proof}

\noindent It is important to be careful about the evaluation of the Fourier coefficients. If these can only be computed using numerical integration, then we have essentially shifted the computational load, but not reduced the overall complexity. For many common functions of interest, such as the mean ($f(x) = x$), second moment ($f(x) = x^2$) and indeed any higher-order moments, then it is easy to see that the Fourier coefficients can be calculated symbolically, and so concerns on this front can be allayed. Furthermore, if there are functions whose Fourier coefficients cannot be found symbolically, but nevertheless are commonly-used, then it may be reasonable to assume that these have been pre-computed to high accuracy and stored, and thus should not count against the complexity of any individual run. However, more generally the Fourier coefficients should be symbolically computable in order that the method attains the advertised speed-up once all of the complexity is counted.

\section{QMCI of the Product of two Correlated Random Variables}
\label{corr}

\noindent As the proposed method of QMCI applies to quantum states that encode \textit{multivariate} probability distributions, a natural next step is to apply the proposed method to functions of multiple random variables. For the sake of definiteness, and for simplicity, we restrict the following analysis to computing the product of functions of two random variables. Moreover, this setting covers computation of correlation and covariance, and indeed most quantities that may be of interest in practice. However, in principle the following approach of using trigonometric identities to break the Monte Carlo integral down into terms that can estimated using QAE also applies to the product of (functions of) three or more random variables.

\begin{corollary}
The quantity, $\mu = \mathbb{E}(f(X)$ $g(Y))$, where $(X , Y) \sim p(x^{(i)}, x^{(j)})$ and $f$ and $g$ are functions that are continuous in value and first derivative and whose second and third derivatives are piecewise-continuous and bounded, can be estimated with MSE $\in \Theta(q^{-\lambda})$, where $q$ is the number of uses of a circuit preparing the quantum state $\ket{p}$ (ie, as defined in (\ref{eqn10})) and $\lambda$ is the convergence rate of some QAE subroutine (ie, as defined in (\ref{eq50})) which operates on circuits of the form defined in Fig.~\ref{fig1}.
\end{corollary}

\begin{proof}[Proof (sketch).]  The idea is to build periodic piecewise functions for both $f$ and $g$, and thus expand each as a Fourier series. This results in a double sum each term of which contains products of sines and cosines. In particular, these will always be either $\cos(\rho)\cos(\sigma)$, $\sin(\rho)\sin(\sigma)$ or $\cos(\rho) \sin(\sigma)$ for some $\rho, \, \sigma$. The next step is to use trigonometric identities to express these terms as functions of single sines and cosines (with correspondingly altered angles). Thus we can decompose the sum representing the expectation we need to compute into a double sum over sines and cosines. The analysis then follows along the same lines as that of Theorem~1, and so to avoid tedious repetition we defer the full proof to Appendix~\ref{appA}.
\end{proof}

\section{Numerical Results}
\label{num}

\begin{figure}[!t]
\centering
\includegraphics[width = \linewidth]{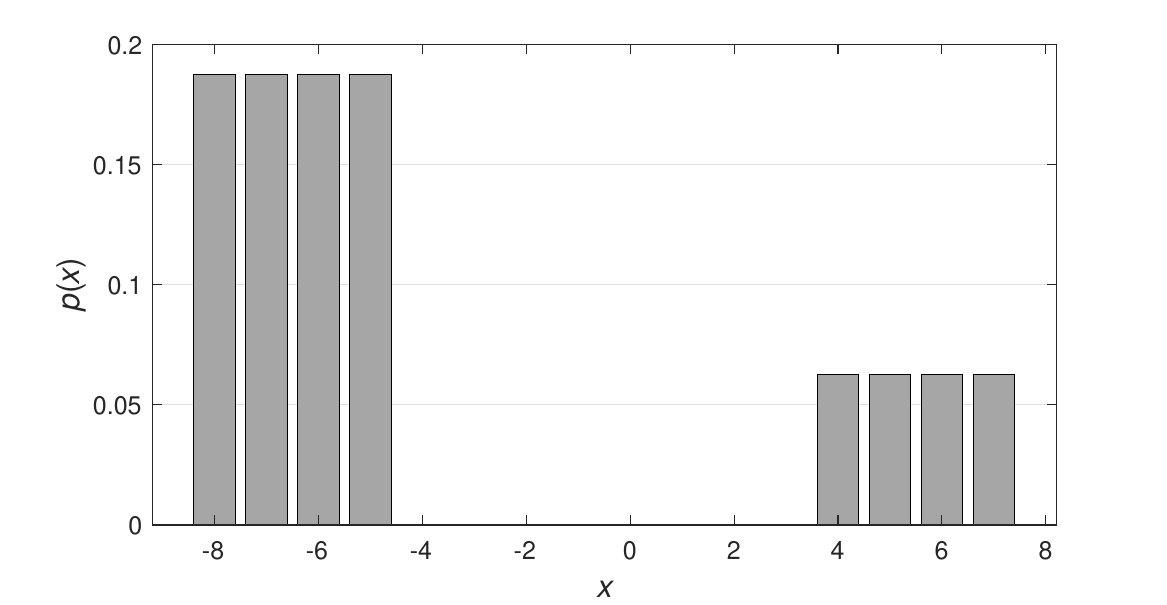} \\
(a)\\
\includegraphics[width = 0.5\linewidth]{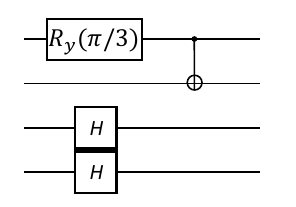}\\
(b) \\
\captionsetup{width=0.95\linewidth}
\caption{(a) The probability distribution used for the numerical results; (b) a circuit that prepares a quantum state encoding this distribution.}
\label{fig04}
\end{figure}

\begin{figure}[!t]
\centering
\includegraphics[width = \linewidth]{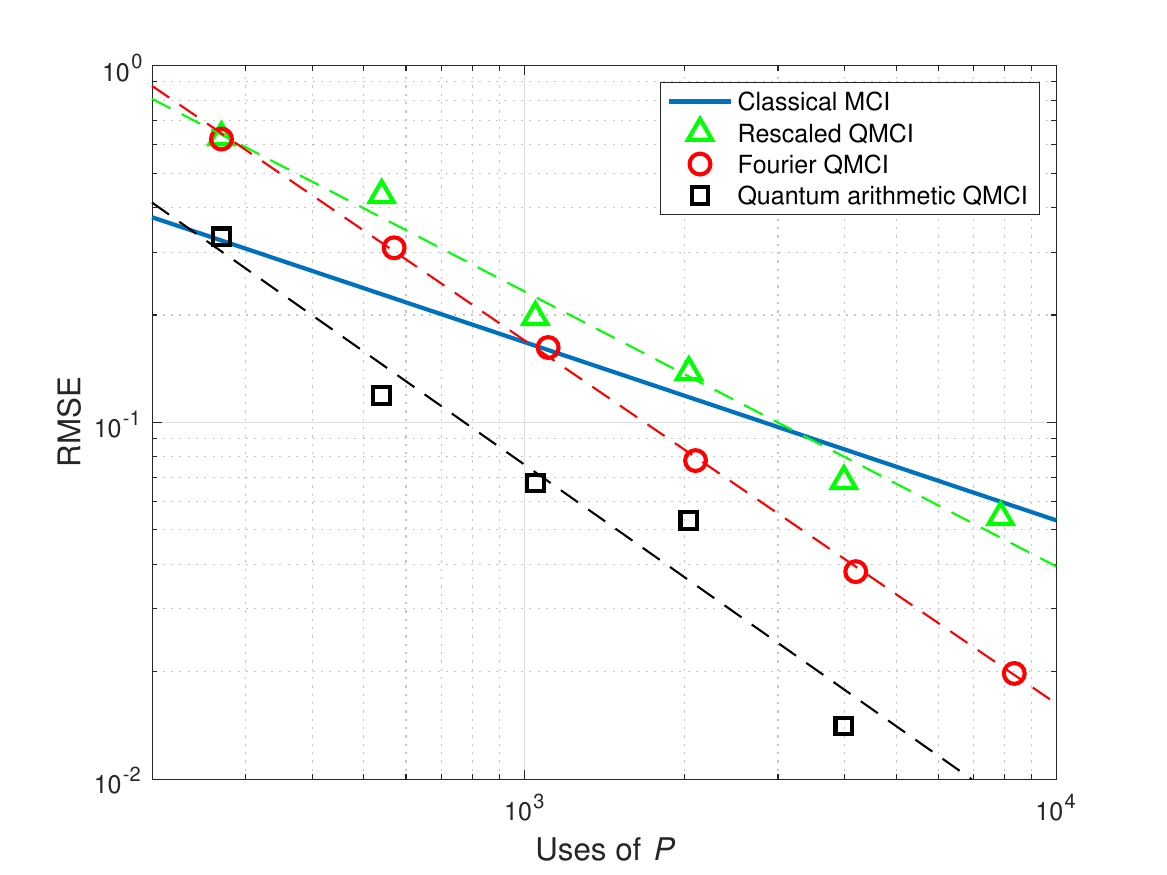}
\captionsetup{width=0.95\linewidth}
\caption{A comparison of classical MCI; rescaled QMCI; Fourier QMCI and conventional QMCI (with quantum arithmetic) for the probability distribution shown in Fig.~\ref{fig04}(a).}
\label{fig05}
\end{figure}

\noindent To illustrate the practical utility of our method of QMCI, we present some simulated results. So that we can compare to classical MCI the rescaled QMCI method of Stamatopoulos \textit{et al} \cite{Stamatopoulos_2020} and conventional QMCI (with quantum arithmetic, but still taking advantage of QPE-free forms of QAE) we consider the case where $f(x) = x$, that is we simply compute the mean of the sampled distribution without applying any other function. We compare for a 16-point univariate probability distribution supported on $\{ -8, -7, \dots, 7 \}$, shown in Fig.~\ref{fig04}(a), which is encoded in a quantum state that can be prepared using the circuit shown in Fig.\ref{fig04}(b). This choice of probability distribution and state preparation circuit has been made largely for practical reasons: there is little to be gained (in terms of proving the principle) by addressing a multivariate distribution, and this would significantly slow down the simulations; and on a similar note, it is desirable to use some probability distribution whose encoding state can be prepared in a very shallow circuit. We have avoided a distribution whose mean is the median of the support, as in this case the rescaled QMCI method can perform artificially well.\\
\indent The results are shown in Fig.~\ref{fig05}, where the simulated results for ``Fourier QMCI'' (the method we propose in this paper), ``Rescaled QMCI'' (the method of Stamatopoulos \textit{et al}) and ``Quantum arithmetic QMCI'' (conventional QMCI using arithmetic subcircuits) were averaged over 500 runs. We can see that for sufficiently small desired root-MSE (RMSE), Fourier QMCI outperforms both Classical MCI and Rescaled QMCI, as expected. Indeed, the slope of the line-of-best-fit for Fourier QMCI is $-1.02$, which is very close to the theoretical value of $-1$ (note that this is \textit{root}-MSE, hence $-1$ not $-2$ as for MSE), and for rescaled QMCI is $-0.771$, clearly inferior and not too far off the theoretical value of $0.667$. It is also worth highlighting that the ``cross-over'' point between Fourier QMCI and classical MCI occurs very close to the third data-point of the former. This represents $1100$ uses of $P$ in total, but more significantly a circuit depth of only 8 sequential uses of $Q$ in any single quantum circuit. 

Finally, we can compare to conventional QMCI (with quantum arithmetic) -- as expected, this performs a little better than Fourier QMCI, although most significantly each has the same rate of decay of RMSE with number of uses of $P$ (the line of best fit for ``Quantum arithmetic QMCI'' is $-1.05 \approx -1$). It is however, worth remarking on the discrepancy between the constant factors for each. The lines of best fit have equations:
\begin{align*}
\text{RMSE(Fourier QMCI)} & = 194 q^{-1.02} \\
\text{RMSE(Quantum arithmetic QMCI)} & = 107 q^{-1.05} \\
\end{align*}
from which we can see that the constant factor is about twice as large for Fourier QMCI as conventional QMCI ($194/107 = 1.81$), which means that to achieve the desired RMSE the deepest circuits will require twice as many Grover iterates for Fourier QMCI compared to conventional QMCI.

However, this will generally still translate into a significant circuit depth reduction for Fourier QMCI when the complexity of the quantum arithmetic is taken into account. In order to construct a suitable circuit $R$ for the function $f(x) = x$ first the bank of rotation gates would be needed to prepare $\sin (x)$ (for instance -- or some other trigonometric function), and then arithmetic circuits for the functions arcsine and square root would be needed. Ref.~\cite{arithref} provides optimised circuits for each of these, and for example we can see from Ref.~\cite[Table II]{arithref} that implementing arcsine in quantum arithmetic would require more than 100 additional qubits and nearly 5000 Toffoli gates, even in the least accurate case considered. Plugging in values to the various expressions for Toffoli counts in Ref.~\cite{arithref} suggests that, even were the required accuracy to be relaxed further, the number of additional Toffolis would still run into the 1000s. Thus we can see that doubling the number of Grover iterates to achieve the desired RMSE (ie, for Fourier QMCI compared to conventional QMC) is a relatively modest price to pay to avoid the heavy cost of quantum arithmetic.

\section{Discussion}
\label{disc}

\noindent In this paper we have proposed a method for performing QMCI by expanding the Monte Carlo integral as a Fourier series and estimating each component individually using amplitude estimation. Our primary motivation for this was to achieve the full quadratic quantum advantage without incurring the need to perform arithmetic operations on the quantum computer, however an added bonus is that our method of QMCI has an innate flexibility that allows it to be readily adapted to a wide variety of settings of real-world relevance. Specifically, we have shown that QMCI can be performed when any function is to be applied to the random samples (with only a mild technical restriction on the smoothness of that function). Moreover, even the expectation of functions of multiple variables can be computed -- allowing the quadratic advantage to be achieved when estimating, for example, the covariance and correlation of some data.

\subsection*{\hspace{-0.55cm} Will Fourier QMCI yield a useful quantum advantage in the NISQ era?}

To answer this question, we first must clarify what, exactly, we mean by ``the NISQ era''. It is fast becoming clear that the dichotomy between the NISQ era and the Full-Scale Fault-Tolerant (FSFT) era is too simplistic -- at least when viewing NISQs in terms of their original definition by Preskill \cite{Preskill_2018}. In these strict terms, the quantum community has very much been focussed on variational algorithms as the only route to NISQ era advantage, however this disposition is somewhat myopic. If the predictions in scaling-up of qubit number (if not fidelity) that hardware manufacturers are making (see eg. Ref \cite{IBM}) come to fruition, then it is reasonable to look forward to a time when some qubits will be allocated as ancillas to enable some degree of error-correction (if not full fault-tolerance) and others as ancillas to perform entanglement distribution to reduce the depth of otherwise costly qubit routing subcircuits (see eg. Ref. \cite{Beaudrap_2020, HdBpatent}), thus effectively facilitating the simulation of a quantum computer with lower qubit-number but higher quantum-volume on the hardware. In such a case, it will in turn be reasonable to think not only of running quantum algorithms with variationally-optimised ans\"{a}tze, but also those with explicitly-constructed circuits, such as QAE, so long as they can adequately handle the inclement noise. Removing the need to perform depth-consuming arithmetic operations on the quantum computer, as we have done in this paper, is an important step towards this.

\subsection*{\hspace{-0.55cm} Will Fourier QMCI still be useful beyond the NISQ era?}

One of the most enduring ideas in quantum algorithm design is that, because quantum computers will not offer useful computational speed-ups for all problems, the focus should therefore be on using the quantum computer only when it offers a computational advantage. Moreover, even within a single task those parts of the computation that can be handed off to a classical computer without adversely impacting overall performance should be. This essential paradigm is unlikely to be restricted solely to the NISQ era, especially as the transition from the NISQ era to FSFT era is unlikely to be particularly sharp (as outlined above). In this paper, we have shown how the task of Monte Carlo integration can be sped-up by using a quantum computer to perform amplitude estimation (where there is a quantum advantage), but not arithmetic (where there is no quantum advantage). We therefore expect that the techniques we have proposed will find useful application for the foreseeable future.

\section*{\hspace{-0.55cm} Acknowledgement}

\noindent Thanks to Roland Guichard for writing the code used for the numerical results, and to Ross Duncan for reviewing the paper. Also, special thanks to Zen Harper for pointing out the additional requirement of piecewise-continuity and boundedness on the second and third derivatives that I had implicitly assumed in the first version, and for providing the more rigorous proof of the Fourier series convergence.\\



\balance

\onecolumn
\pagebreak

\appendix

\section{Proof of Corollary 4}
\label{appA}

\vspace{0.5cm}

\setcounter{thm}{3}

\begin{corollary}
The quantity, $\mu = \mathbb{E}(f(X)$ $g(Y))$, where $(X , Y) \sim p(x^{(i)}, x^{(j)})$ and $f$ and $g$ are functions that are continuous in value and first derivative and whose second and third derivatives are piecewise-continuous and bounded, can be estimated with MSE $\in \Theta(q^{-\lambda})$, where $q$ is the number of uses of a circuit preparing the quantum state $\ket{p}$ (ie, as defined in (\ref{eqn10})) and $\lambda$ is the convergence rate of some QAE subroutine (ie, as defined in (\ref{eq50})) which operates on circuits of the form defined in Fig.~\ref{fig1}.
\end{corollary}

\begin{proof}

The proof follows similar lines to the univariate case. First we construct periodic piecewise functions to extend the two functions of interest, then show that it is possible to truncate and estimate each Fourier component individually to achieve the claimed convergence. To start with, we let $\mathrm{f}$ and $\mathrm{g}$ be periodic piecewise functions (continuous in value and first derivative, and whose second and third derivatives are piecewise-continuous and bounded) for $f$ and $g$ respectively -- so that $\mathrm{f}(x) = f(x)$ whenever $p(x)$ is non-zero (and likewise for $\mathrm{g}(x)$). Thus we can write the Fourier series:
\begin{align}
\mathrm{f}(x) & = c^{(1)} + \sum_{n=1}^{\infty} \frac{1}{n^3} \left( \tilde{a}^{(1)}_n \cos (n \omega^{(1)} x) + \tilde{b}^{(1)}_n \sin (n \omega^{(1)} x) \right) \\
\mathrm{g}(x) & = c^{(2)} + \sum_{m=1}^{\infty} \frac{1}{m^3} \left( \tilde{a}^{(2)}_m \cos (m \omega^{(2)} x) + \tilde{b}^{(2)}_m \sin (m \omega^{(2)} x) \right)
\end{align}
where $ |\tilde{a}^{(1)}_n| , \, |\, \tilde{a}^{(2)}_m|, \, |\tilde{b}^{(1)}_n|, \, |\tilde{b}^{(2)}_m| \in \mathcal{O}(1)$. Thus we can write the quantity of interest, $\mu$, in terms of the Fourier decompositions:
\begin{align}
\mu =&  \mathbb{E}(f(X)g(Y)) \nonumber \\
= &  \sum_{x^{(i)}, x^{(j)}} \left( c^{(1)} + \sum_{n=1}^{\infty} \frac{1}{n^3} \left( \tilde{a}^{(1)}_n \cos (n \omega^{(1)} x^{(i)}) + \tilde{b}^{(1)}_n \sin (n \omega^{(1)} x^{(i)}) \right) \right) \nonumber \\
& \,\,\,\,\,\,\,\,\,\,\,\,\,\,\,\,\,\,\,\,\,\,\,\,\,\,\,\,\,\,\,\, \left( c^{(2)} + \sum_{m=1}^{\infty} \frac{1}{m^3} \left( \tilde{a}^{(2)}_m \cos (m \omega^{(2)} x^{(j)}) + \tilde{b}^{(2)}_m \sin (m \omega^{(2)} x^{(j)}) \right) \right) p(x^{(i)}, x^{(j)}) \nonumber \\
& = c^{(1)} + c^{(2)} + \sum_{n=1}^{\infty} \sum_{m=1}^{\infty} \frac{1}{(nm)^3}  \Bigg(  \sum_{x^{(i)}, x^{(j)}} \left( \tilde{a}^{(1)}_n \cos (n \omega^{(1)} x^{(i)}) + \tilde{b}^{(1)}_n \sin (n \omega^{(1)} x^{(i)}) \right) \nonumber \\
& \,\,\,\,\,\,\,\,\,\,\,\,\,\,\,\,\,\,\,\,\,\,\,\,\,\,\,\,\,\,\,\,\,\,\,\,\,\,\,\,\,\,\,\,\,\,\,\,\,\,\,\,\,\,\,\,\,\,\,\,\,\,\,\, \,\,\,\,\,\,\,\,\,\,\,\,\,\,\,\,\,\,\,\,\,\,\,\,\,\,\,\,\,\,\,\, \left( \tilde{a}^{(2)}_m \cos (m \omega^{(2)} x^{(j)}) + \tilde{b}^{(2)}_m \sin (m \omega^{(2)} x^{(j)}) \right)  p(x^{(i)}, x^{(j)}) \Bigg)
\end{align}
Once again, we assume that the infinite sums are truncated at $\lceil q_0^{\lambda / 4} \rceil$, thus we have:
\begin{align}
\mu & = c^{(1)} + c^{(2)} + \sum_{n=1}^{\lceil q_0^{\lambda / 4} \rceil} \sum_{m=1}^{\lceil q_0^{\lambda / 4} \rceil} \frac{1}{(nm)^3}  \Bigg(  \sum_{x^{(i)}, x^{(j)}} \left( \tilde{a}^{(1)}_n \cos (n \omega^{(1)} x^{(i)}) + \tilde{b}^{(1)}_n \sin (n \omega^{(1)} x^{(i)}) \right) \nonumber \\
\label{app1000}
& \,\,\,\,\,\,\,\,\,\,\,\,\,\,\,\,\,\,\,\,\,\,\,\,\,\,\,\,\,\,\,\,\,\,\,\,\,\,\,\,\,\,\,\,\,\,\,\,\,\,\,\,\,\,\,\,\,\,\,\,\,\,\,\, \,\,\,\,\,\,\,\,\,\,\,\,\,\,\,\,\,\,\,\,\,\,\,\,\,\,\,\,\,\,\,\, \left( \tilde{a}^{(2)}_m \cos (m \omega^{(2)} x^{(j)}) + \tilde{b}^{(2)}_m \sin (m \omega^{(2)} x^{(j)}) \right)  p(x^{(i)}, x^{(j)}) \Bigg) + \eta
\end{align}
where
\begin{align}
\eta = & \sum_{n=\lceil q_0^{\lambda / 4} \rceil + 1}^{\infty} \,\, \sum_{m=1}^{\infty} \frac{1}{(nm)^3}  \Bigg(  \sum_{x^{(i)}, x^{(j)}} \left( \tilde{a}^{(1)}_n \cos (n \omega^{(1)} x^{(i)}) + \tilde{b}^{(1)}_n \sin (n \omega^{(1)} x^{(i)}) \right) \nonumber \\
& \,\,\,\,\,\,\,\,\,\,\,\,\,\,\,\,\,\,\,\,\,\,\,\,\,\,\,\,\,\,\,\,\,\,\,\,\,\,\,\,\,\,\,\,\,\,\,\,\,\,\,\,\,\,\,\,\,\,\,\,\,\,\,\, \,\,\,\,\,\,\,\,\,\,\,\,\,\,\,\,\,\,\,\,\,\,\,\,\,\,\,\,\,\,\,\, \left( \tilde{a}^{(2)}_m \cos (m \omega^{(2)} x^{(j)}) + \tilde{b}^{(2)}_m \sin (m \omega^{(2)} x^{(j)}) \right)  p(x^{(i)}, x^{(j)}) \Bigg) \nonumber \\
& \,\,\,\,\,\,\, + \sum_{n=1}^{\infty}  \,\, \sum_{m=\lceil q_0^{\lambda / 4} \rceil+1}^{\infty} \frac{1}{(nm)^3}  \Bigg(  \sum_{x^{(i)}, x^{(j)}} \left( \tilde{a}^{(1)}_n \cos (n \omega^{(1)} x^{(i)}) + \tilde{b}^{(1)}_n \sin (n \omega^{(1)} x^{(i)}) \right) \nonumber \\
& \,\,\,\,\,\,\,\,\,\,\,\,\,\,\,\,\,\,\,\,\,\,\,\,\,\,\,\,\,\,\,\,\,\,\,\,\,\,\,\,\,\,\,\,\,\,\,\,\,\,\,\,\,\,\,\,\,\,\,\,\,\,\,\,\,\,\,\,\,\,\, \,\,\,\,\,\,\,\,\,\,\,\,\,\,\,\,\,\,\,\,\,\,\,\,\,\,\,\,\,\,\,\, \left( \tilde{a}^{(2)}_m \cos (m \omega^{(2)} x^{(j)}) + \tilde{b}^{(2)}_m \sin (m \omega^{(2)} x^{(j)}) \right)  p(x^{(i)}, x^{(j)}) \Bigg) \nonumber \\
&  \,\,\,\,\,\,\, - \sum_{n=\lceil q_0^{\lambda / 4} \rceil+1}^{\infty}  \,\, \sum_{m=\lceil q_0^{\lambda / 4} \rceil+1}^{\infty} \frac{1}{(nm)^3}  \Bigg(  \sum_{x^{(i)}, x^{(j)}} \left( \tilde{a}^{(1)}_n \cos (n \omega^{(1)} x^{(i)}) + \tilde{b}^{(1)}_n \sin (n \omega^{(1)} x^{(i)}) \right) \nonumber \\
& \,\,\,\,\,\,\,\,\,\,\,\,\,\,\,\,\,\,\,\,\,\,\,\,\,\,\,\,\,\,\,\,\,\,\,\,\,\,\,\,\,\,\,\,\,\,\,\,\,\,\,\,\,\,\,\,\,\,\,\,\,\,\,\,\,\,\,\,\,\,\, \,\,\,\,\,\,\,\,\,\,\,\,\,\,\,\,\,\,\,\,\,\,\,\,\,\,\,\,\,\,\,\, \left( \tilde{a}^{(2)}_m \cos (m \omega^{(2)} x^{(j)}) + \tilde{b}^{(2)}_m \sin (m \omega^{(2)} x^{(j)}) \right)  p(x^{(i)}, x^{(j)}) \Bigg) \nonumber \\
\implies | \eta | \leq & \Bigg| \sum_{n=\lceil q_0^{\lambda / 4} \rceil + 1}^{\infty}  \,\, \sum_{m=1}^{\infty} \frac{1}{(nm)^3}  \Bigg(  \sum_{x^{(i)}, x^{(j)}} \left( \tilde{a}^{(1)}_n \cos (n \omega^{(1)} x^{(i)}) + \tilde{b}^{(1)}_n \sin (n \omega^{(1)} x^{(i)}) \right) \nonumber \\
& \,\,\,\,\,\,\,\,\,\,\,\,\,\,\,\,\,\,\,\,\,\,\,\,\,\,\,\,\,\,\,\,\,\,\,\,\,\,\,\,\,\,\,\,\,\,\,\,\,\,\,\,\,\,\,\,\,\,\,\,\,\,\,\, \,\,\,\,\,\,\,\,\,\,\,\,\,\,\,\,\,\,\,\,\,\,\,\,\,\,\,\,\,\,\,\, \left( \tilde{a}^{(2)}_m \cos (m \omega^{(2)} x^{(j)}) + \tilde{b}^{(2)}_m \sin (m \omega^{(2)} x^{(j)}) \right)  p(x^{(i)}, x^{(j)}) \Bigg) \Bigg| \nonumber \\
& \,\,\,\,\,\,\, + \Bigg| \sum_{n=1}^{\infty}  \,\, \sum_{m=\lceil q_0^{\lambda / 4}  \rceil + 1}^{\infty} \frac{1}{(nm)^3}  \Bigg(  \sum_{x^{(i)}, x^{(j)}} \left( \tilde{a}^{(1)}_n \cos (n \omega^{(1)} x^{(i)}) + \tilde{b}^{(1)}_n \sin (n \omega^{(1)} x^{(i)}) \right)  \nonumber \\
& \,\,\,\,\,\,\,\,\,\,\,\,\,\,\,\,\,\,\,\,\,\,\,\,\,\,\,\,\,\,\,\,\,\,\,\,\,\,\,\,\,\,\,\,\,\,\,\,\,\,\,\,\,\,\,\,\,\,\,\,\,\,\,\,\,\,\,\,\,\,\, \,\,\,\,\,\,\,\,\,\,\,\,\,\,\,\,\,\,\,\,\,\,\,\,\,\,\,\,\,\,\,\, \left( \tilde{a}^{(2)}_m \cos (m \omega^{(2)} x^{(j)}) + \tilde{b}^{(2)}_m \sin (m \omega^{(2)} x^{(j)}) \right)  p(x^{(i)}, x^{(j)}) \Bigg) \Bigg| \nonumber \\
 & \,\,\,\,\,\,\, + \Bigg| \sum_{n=\lceil q_0^{\lambda / 4} \rceil+1}^{\infty}  \,\, \sum_{m=\lceil q_0^{\lambda / 4} \rceil+1}^{\infty} \frac{1}{(nm)^3}  \Bigg(  \sum_{x^{(i)}, x^{(j)}} \left( \tilde{a}^{(1)}_n \cos (n \omega^{(1)} x^{(i)}) + \tilde{b}^{(1)}_n \sin (n \omega^{(1)} x^{(i)}) \right) \nonumber \\
& \,\,\,\,\,\,\,\,\,\,\,\,\,\,\,\,\,\,\,\,\,\,\,\,\,\,\,\,\,\,\,\,\,\,\,\,\,\,\,\,\,\,\,\,\,\,\,\,\,\,\,\,\,\,\,\,\,\,\,\,\,\,\,\,\,\,\,\,\,\,\, \,\,\,\,\,\,\,\,\,\,\,\,\,\,\,\,\,\,\,\,\,\,\,\,\,\,\,\,\,\,\,\, \left( \tilde{a}^{(2)}_m \cos (m \omega^{(2)} x^{(j)}) + \tilde{b}^{(2)}_m \sin (m \omega^{(2)} x^{(j)}) \right)  p(x^{(i)}, x^{(j)}) \Bigg) \Bigg| \nonumber \\
\leq &  \sum_{n=\lceil q_0^{\lambda / 4} \rceil + 1}^{\infty }  \,\, \sum_{m=1}^{\infty} \frac{1}{(nm)^3} 4 k_2^2 + \sum_{n=1}^{\infty}  \,\, \sum_{m=\lceil q_0^{\lambda / 4} \rceil + 1}^{\infty} \frac{1}{(nm)^3}  4 k_2^2 + \sum_{n=\lceil q_0^{\lambda / 4} \rceil + 1}^{\infty}  \,\, \sum_{m=\lceil q_0^{\lambda / 4} \rceil + 1}^{\infty} \frac{1}{(nm)^3}  4 k_2^2 \nonumber \\
\leq &  12  k_2^2 \sum_{n=\lceil q_0^{\lambda / 4} \rceil + 1}^{\infty }  \,\, \sum_{m=1}^{\infty} \frac{1}{(nm)^3}  \nonumber \\
\leq & k_5 \left( \lceil q_0^{\lambda / 4} \rceil \right)^{-2}  \nonumber \\
\label{newapp10}
\leq &  k_6 q_0^{-\lambda / 2} 
\end{align}

\pagebreak

We now address the other terms in (\ref{app1000}):
\begin{align}
  \sum_{x^{(i)}, x^{(j)}} & \left( \tilde{a}^{(1)}_n \cos (n \omega^{(1)} x^{(i)}) + \tilde{b}^{(1)}_n \sin (n \omega^{(1)} x^{(i)}) \right) \left( \tilde{a}^{(2)}_m \cos (m \omega^{(2)} x^{(j)}) + \tilde{b}^{(2)}_m \sin (m \omega^{(2)} x^{(j)}) \right)  p(x^{(i)}, x^{(j)})   \nonumber \\
& =  \tilde{a}^{(1)}_n   \tilde{a}^{(2)}_m  \sum_{x^{(i)}, x^{(j)}}  \cos (n \omega^{(1)} x^{(i)})\cos (m \omega^{(2)} x^{(j)}) p(x^{(i)}, x^{(j)}) \nonumber \\
& \,\,\,\,\,  \,\,\,\,\,   \,\,\,\,\, + \tilde{a}^{(1)}_n   \tilde{b}^{(2)}_m  \sum_{x^{(i)}, x^{(j)}} \cos (n \omega^{(1)} x^{(i)})\sin (m \omega^{(2)} x^{(j)}) p(x^{(i)}, x^{(j)}) \nonumber \\
& \,\,\,\,\,  \,\,\,\,\,   \,\,\,\,\,  + \tilde{b}^{(1)}_n   \tilde{a}^{(2)}_m  \sum_{x^{(i)}, x^{(j)}} \sin (n \omega^{(1)} x^{(i)})\cos (m \omega^{(2)} x^{(j)}) p(x^{(i)}, x^{(j)}) \nonumber \\
& \,\,\,\,\,  \,\,\,\,\,   \,\,\,\,\, + \tilde{b}^{(1)}_n   \tilde{b}^{(2)}_m  \sum_{x^{(i)}, x^{(j)}} \sin (n \omega^{(1)} x^{(i)})\sin (m \omega^{(2)} x^{(j)}) p(x^{(i)}, x^{(j)}) \nonumber \\
& =  0.5 \tilde{a}^{(1)}_n  \tilde{a}^{(2)}_m  \sum_{x^{(i)}, x^{(j)}}  \cos (n \omega^{(1)} x^{(i)} - m \omega^{(2)} x^{(j)}) p(x^{(i)}, x^{(j)})   \nonumber \\
& \,\,\,\,\,  \,\,\,\,\,   \,\,\,\,\,   \,\,\,\,\,  \,\,\,\,\,   \,\,\,\,\, +  0.5 \tilde{a}^{(1)}_n  \tilde{a}^{(2)}_m  \sum_{x^{(i)}, x^{(j)}}  \cos (n \omega^{(1)} x^{(i)} + m \omega^{(2)} x^{(j)}) p(x^{(i)}, x^{(j)}) \nonumber \\
& \,\,\,\,\,  \,\,\,\,\,   \,\,\,\,\, +  0.5 \tilde{a}^{(1)}_n  \tilde{b}^{(2)}_m  \sum_{x^{(i)}, x^{(j)}}  \sin (n \omega^{(1)} x^{(i)} + m \omega^{(2)} x^{(j)}) p(x^{(i)}, x^{(j)})   \nonumber \\
& \,\,\,\,\,  \,\,\,\,\,   \,\,\,\,\,   \,\,\,\,\,  \,\,\,\,\,   \,\,\,\,\, -  0.5 \tilde{a}^{(1)}_n  \tilde{b}^{(2)}_m  \sum_{x^{(i)}, x^{(j)}}  \sin (n \omega^{(1)} x^{(i)} - m \omega^{(2)} x^{(j)}) p(x^{(i)}, x^{(j)}) \nonumber \\
& \,\,\,\,\,  \,\,\,\,\,   \,\,\,\,\, +  0.5 \tilde{b}^{(1)}_n  \tilde{a}^{(2)}_m  \sum_{x^{(i)}, x^{(j)}}  \sin (n \omega^{(1)} x^{(i)} + m \omega^{(2)} x^{(j)}) p(x^{(i)}, x^{(j)})   \nonumber \\
& \,\,\,\,\,  \,\,\,\,\,   \,\,\,\,\,   \,\,\,\,\,  \,\,\,\,\,   \,\,\,\,\, + 0.5 \tilde{b}^{(1)}_n  \tilde{a}^{(2)}_m  \sum_{x^{(i)}, x^{(j)}}  \sin (n \omega^{(1)} x^{(i)} - m \omega^{(2)} x^{(j)}) p(x^{(i)}, x^{(j)}) \nonumber \\
& \,\,\,\,\,  \,\,\,\,\,   \,\,\,\,\, + 0.5 \tilde{b}^{(1)}_n  \tilde{b}^{(2)}_m  \sum_{x^{(i)}, x^{(j)}}  \cos (n \omega^{(1)} x^{(i)} - m \omega^{(2)} x^{(j)}) p(x^{(i)}, x^{(j)})   \nonumber \\
& \,\,\,\,\,  \,\,\,\,\,   \,\,\,\,\,   \,\,\,\,\,  \,\,\,\,\,   \,\,\,\,\, -  0.5 \tilde{b}^{(1)}_n  \tilde{b}^{(2)}_m  \sum_{x^{(i)}, x^{(j)}}  \cos (n \omega^{(1)} x^{(i)} + m \omega^{(2)} x^{(j)}) p(x^{(i)}, x^{(j)}) \nonumber \\
& =  0.5 ( \tilde{a}^{(1)}_n  \tilde{a}^{(2)}_m  + \tilde{b}^{(1)}_n  \tilde{b}^{(2)}_m ) \sum_{x^{(i)}, x^{(j)}}  \cos (n \omega^{(1)} x^{(i)} - m \omega^{(2)} x^{(j)}) p(x^{(i)}, x^{(j)}) \nonumber \\
& \,\,\,\,\,  \,\,\,\,\,   \,\,\,\,\, + 0.5 (\tilde{a}^{(1)}_n  \tilde{a}^{(2)}_m - \tilde{b}^{(1)}_n  \tilde{b}^{(2)}_m ) \sum_{x^{(i)}, x^{(j)}}  \cos (n \omega^{(1)} x^{(i)} + m \omega^{(2)} x^{(j)}) p(x^{(i)}, x^{(j)}) \nonumber \\
&\,\,\,\,\,  \,\,\,\,\,   \,\,\,\,\, + 0.5( \tilde{a}^{(1)}_n  \tilde{b}^{(2)}_m + \tilde{b}^{(1)}_n  \tilde{a}^{(2)}_m ) \sum_{x^{(i)}, x^{(j)}}  \sin (n \omega^{(1)} x^{(i)} + m \omega^{(2)} x^{(j)}) p(x^{(i)} , x^{(j)}) \nonumber \\
\label{eqn5000}
& \,\,\,\,\,  \,\,\,\,\,   \,\,\,\,\, + 0.5( \tilde{a}^{(1)}_n  \tilde{b}^{(2)}_m - \tilde{b}^{(1)}_n  \tilde{a}^{(2)}_m ) \sum_{x^{(i)}, x^{(j)}}  \sin (n \omega^{(1)} x^{(i)} - m \omega^{(2)} x^{(j)}) p(x^{(i)} , x^{(j)})
\end{align}

\pagebreak

We can now use QAE to find an approximation of each of the four terms in the final line of (\ref{eqn5000}). To do this, we define:
\begin{align}
\mu_{nm}^{(a)} & = \sum_{x^{(i)}, x^{(j)}}  \cos (n \omega^{(1)} x^{(i)} - m \omega^{(2)} x^{(j)}) p(x^{(i)}, x^{(j)}) \\
\mu_{nm}^{(b)} & = \sum_{x^{(i)}, x^{(j)}}  \cos (n \omega^{(1)} x^{(i)} + m \omega^{(2)} x^{(j)}) p(x^{(i)}, x^{(j)}) \\
\mu_{nm}^{(c)} & = \sum_{x^{(i)}, x^{(j)}}  \sin (n \omega^{(1)} x^{(i)} + m \omega^{(2)} x^{(j)}) p(x^{(i)} , x^{(j)}) \\
\mu_{nm}^{(d)} & = \sum_{x^{(i)}, x^{(j)}}  \sin (n \omega^{(1)} x^{(i)} - m \omega^{(2)} x^{(j)}) p(x^{(i)} , x^{(j)})
\end{align}
These can each be estimated using QAE. In particular we define a family of parameterised circuits, $A'(P,i,j,\beta,$ $n,m, \omega^{(1)}, \omega^{(2)})$ shown in Fig.~\ref{fig1}, where the support of the $i^{th}$ dimension of the multivariate probability distribution is defined in terms of $x_l^{(i)}$ and $\Delta^{(i)}$ (and analogously for the $j^{th}$ dimension). Once again $\beta$ is set to $\pi / 2$ when the sinusoidal components are being estimated, and in the case of $\mu_{nm}^{(a)}$ and $\mu_{nm}^{(d)}$ we set $m$ to $-m$ to achieve the negation. We let each term have $q_n = q_0 (nm)^{2- \delta}$ uses of $P$ (for some $0 < \delta < 1$), and thus achieve the estimates:
\begin{align}
\mu_{nm}^{(a)} & = \hat{\mu}_{nm}^{(a)} + \nu_{nm}^{(a)} + \epsilon_{nm}^{(a)} \\
\mu_{nm}^{(b)} & = \hat{\mu}_{nm}^{(b)} + \nu_{nm}^{(b)} + \epsilon_{nm}^{(b)} \\
\mu_{nm}^{(c)} & = \hat{\mu}_{nm}^{(c)} + \nu_{nm}^{(c)} + \epsilon_{nm}^{(c)} \\
\mu_{nm}^{(d)} & = \hat{\mu}_{nm}^{(d)} + \nu_{nm}^{(d)} + \epsilon_{nm}^{(d)} 
\end{align}
where again $\hat{\mu}_{nm}^{(a)} $ is the estimate, $ \nu_{nm}^{(a)}$ is the bias of the estimator and $\epsilon_{nm}^{(a)}$ is the zero-mean error of the estimator (and likewise for the other terms). 

\begin{figure}[!t]
\centering
\includegraphics[width = 0.6 \textwidth]{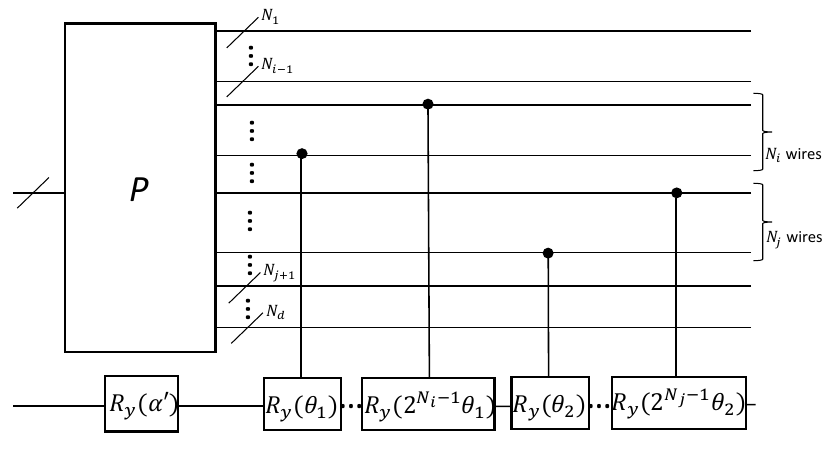}
\captionsetup{width=0.95\linewidth}
\caption{Circuit diagram of $A'(P,i,j,\beta,n,m, \omega^{(1)}, \omega^{(2)})$ where: $\alpha' = n \omega^{(1)} x_l^{(i)} + m \omega^{(2)} x_l^{(j)} - \beta$, $\theta_1 = n \omega^{(1)} \Delta^{(i)}$ and $\theta_2 = m \omega^{(2)} \Delta^{(j)}$.}
\label{fig1}
\end{figure}

Making the appropriate substitutions in (\ref{app1000}) we get
\begin{align}
\mu - \hat{\mu} = 0.5  \sum_{n=1}^{\lceil q_0^{\lambda / 4} \rceil} \sum_{m=1}^{\lceil q_0^{\lambda / 4} \rceil} &  \frac{1}{(nm)^3} \Bigg(  ( \tilde{a}^{(1)}_n  \tilde{a}^{(2)}_m  + \tilde{b}^{(1)}_n  \tilde{b}^{(2)}_m ) (\nu_{nm}^{(a)} + \epsilon_{nm}^{(a)} ) + (\tilde{a}^{(1)}_n  \tilde{a}^{(2)}_m - \tilde{b}^{(1)}_n  \tilde{b}^{(2)}_m ) (\nu_{nm}^{(b)} + \epsilon_{nm}^{(b)} ) \nonumber \\
& + ( \tilde{a}^{(1)}_n  \tilde{b}^{(2)}_m + \tilde{b}^{(1)}_n  \tilde{a}^{(2)}_m )  (\nu_{nm}^{(c)} + \epsilon_{nm}^{(c)} ) + ( \tilde{a}^{(1)}_n  \tilde{b}^{(2)}_m - \tilde{b}^{(1)}_n  \tilde{a}^{(2)}_m ) (\nu_{nm}^{(d)} + \epsilon_{nm}^{(d)} ) \Bigg) + \eta
\end{align}
Once again, using the fact that $ |\tilde{a}^{(1)}_n| , \, |\, \tilde{a}^{(2)}_m|, \, |\tilde{b}^{(1)}_n|, \, |\tilde{b}^{(2)}_m| \in \mathcal{O}(1)$ we use the constant $k_2$ to upper-bound the magnitudes thereof, and get:
\begin{equation}
| \mu - \hat{\mu} | \leq k_2^2 \Bigg| \sum_{n=1}^{\lceil q_0^{\lambda / 4} \rceil} \sum_{m=1}^{\lceil q_0^{\lambda / 4} \rceil} \frac{1}{(nm)^3} \Bigg( \nu_{nm}^{(a)} + \epsilon_{nm}^{(a)} + \nu_{nm}^{(b)} + \epsilon_{nm}^{(b)} + \nu_{nm}^{(c)} + \epsilon_{nm}^{(c)} + \nu_{nm}^{(d)} + \epsilon_{nm}^{(d)} \Bigg) + \eta \Bigg|
\end{equation}
We also define:
\begin{equation}
\nu = k_2^2 \sum_{n=1}^{\lceil q_0^{\lambda / 4} \rceil} \sum_{m=1}^{\lceil q_0^{\lambda / 4} \rceil} \frac{1}{(nm)^3} \nu_{nm}^{(a)} +\nu_{nm}^{(b)} +\nu_{nm}^{(c)} +\nu_{nm}^{(d)} 
\end{equation}
Which thus allows to express:
\begin{align}
\mathbb{E} \left( ( \mu - \hat{\mu} )^2 \right) & \leq \mathbb{E} \left( \Bigg( \nu +  k_2^2 \sum_{n=1}^{\lceil q_0^{\lambda / 4} \rceil}  \sum_{m=1}^{\lceil q_0^{\lambda / 4} \rceil} \frac{1}{(nm)^3} (  \epsilon_{nm}^{(a)} + \epsilon_{nm}^{(b)} + \epsilon_{nm}^{(c)} + \epsilon_{nm}^{(d)} ) + \eta \Bigg)^2 \right) \nonumber \\
\label{app2000}
& = \nu^2 + \eta^2 + 2 \nu \eta + k_2^2 \sum_{n=1}^{\lceil q_0^{\lambda / 4} \rceil}  \sum_{m=1}^{\lceil q_0^{\lambda / 4} \rceil} \frac{1}{(nm)^6} \left(  \mathbb{E}( (\epsilon_{nm}^{(a)})^2) + \mathbb{E}( ( \epsilon_{nm}^{(b)})^2)  +  \mathbb{E}( (  \epsilon_{nm}^{(c)})^2)  + \mathbb{E}( ( \epsilon_{nm}^{(d)})^2)  \right) 
\end{align}
We have already upper-bounded $|\eta | \leq k_6 q_0^{-\lambda /2} $ in (\ref{newapp10}), and we now bound the other terms, recalling that $q_{nm} = q_0 (nm)^{-2 + \delta}$ uses of $P$ for each QAE subroutine. Starting with $\nu$, and noting that $|\nu_{nm}^{(a)}|$ is upper-bounded by the square root of the MSE (and likewise for $|\nu_{nm}^{(b)}|$, $|\nu_{nm}^{(c)}|$ and $|\nu_{nm}^{(d)}|$), we get:
\begin{align}
|\nu| & \leq k_2^2 \sum_{n=1}^{\lceil q_0^{\lambda / 4} \rceil} \sum_{m=1}^{\lceil q_0^{\lambda / 4} \rceil} \frac{4}{(nm)^3} \sqrt{k_1 q_n^{-\lambda}} \nonumber \\
& = 4 k_2^2 \sum_{n=1}^{\lceil q_0^{\lambda / 4} \rceil} \sum_{m=1}^{\lceil q_0^{\lambda / 4} \rceil} \frac{1}{(nm)^3} k_1^{1/2} (q_0 (nm)^{-2 + \delta})^{-\lambda /2}\nonumber \\
& \leq 4 k_1^{1/2} k_2^2 q_0^{-\lambda /2} \sum_{n=1}^{\infty} n^{\lambda(1 -\delta /2) - 3} \sum_{m=1}^{\infty} m^{\lambda(1 -\delta /2) - 3} \nonumber \\
\label{newapp20}
& = k_7 q_0^{-\lambda/2}
\end{align}
by noting that $\lambda \leq 2$. Similarly, again we have that:
\begin{equation}
\mathbb{E}( (\epsilon_{nm}^{(a)})^2), \, \mathbb{E}( (\epsilon_{nm}^{(b)})^2), \, \mathbb{E}( (\epsilon_{nm}^{(c)})^2), \, \mathbb{E}( (\epsilon_{nm}^{(d)})^2) \leq k_1 q_n^{-\lambda}
\end{equation}
with equality for unbiased estimates. Thus we get:
\begin{align}
\sum_{n=1}^{\lceil q_0^{\lambda / 4} \rceil}  \sum_{m=1}^{\lceil q_0^{\lambda / 4} \rceil} \frac{1}{(nm)^6} \left(  \mathbb{E}( (\epsilon_{nm}^{(a)})^2) + \mathbb{E}( ( \epsilon_{nm}^{(b)})^2)  +  \mathbb{E}( (  \epsilon_{nm}^{(c)})^2)  + \mathbb{E}( ( \epsilon_{nm}^{(d)})^2)  \right) & \leq \sum_{n=1}^{\lceil q_0^{\lambda / 4} \rceil}  \sum_{m=1}^{\lceil q_0^{\lambda / 4} \rceil} \frac{1}{(nm)^6} k_1 \left( q_0 (nm)^{-2+\delta} \right)^{-\lambda} \nonumber \\
& = k_1 q_0^{-\lambda} \sum_{n=1}^{\lceil q_0^{\lambda / 4} \rceil} n^{\lambda(2-\delta) - 6}  \sum_{m=1}^{\lceil q_0^{\lambda / 4} \rceil} m^{\lambda(2-\delta) - 6} \nonumber \\
& = k_1 q_0^{-\lambda} \sum_{n=1}^{\infty} n^{\lambda(2-\delta) - 6}  \sum_{m=1}^{\infty} m^{\lambda(2-\delta) - 6} \nonumber \\
\label{newapp30}
& = k_8 q_0^{-\lambda}
\end{align}
again using $\lambda \leq 2$. Substituting (\ref{newapp10}), (\ref{newapp20}) and (\ref{newapp30}) into (\ref{app2000}) we get:
\begin{align}
\mathbb{E} \left( ( \mu - \hat{\mu} )^2 \right) & \leq \left( k_2^2k_8 + (k_6 + k_7)^2 \right) q_0^{-\lambda} \nonumber \\
\label{newapp40}
& \in \mathcal{O}(q_0^{-\lambda})
\end{align}
Even though we haven't shown the lower bound, it is clear that the method presented here won't allow us to do better than $\Omega(q_0^{-\lambda})$, so we can conclude that the complexity is $\Theta(q_0^{-\lambda})$. Finally, we must count up the number of uses of $P$. Once again, we must include the possibility of needing to round up -- we include a total of $(\lceil q_0^{\lambda /4} \rceil)^2$ Fourier components (that is, because both random variables being multiplied have $\lceil q_0^{\lambda /4} \rceil$ terms), and each of these requires four uses of QAE, thus we get that the total number of uses of $P$ can be counted up as:
\begin{align}
q_{tot} & = \sum_{n=1}^{\lceil q_0^{\lambda / 4} \rceil}  \sum_{m=1}^{\lceil q_0^{\lambda / 4} \rceil} 4 k_1 q_0 (nm)^{-2 + \delta} + 4 (\lceil q_0^{\lambda /4} \rceil)^2 \nonumber \\
& \leq 4 k_1 q_0 \sum_{n=1}^{\infty} n^{-2+ \delta} \sum_{m=1}^{\infty} m^{-2+ \delta} + 4 (\lceil q_0^{\lambda /4} \rceil)^2 \nonumber \\
& \in \Theta (q_0)
 \end{align}
 using the fact that $\delta < 1$. Thus we have shown that the total number of uses of $P$ is just a constant multiple of $q_0$, and from (\ref{newapp40}) we have shown that the MSE decays as $q_0^{-\lambda}$, thus proving the Corollary.
\end{proof}

\newpage

\section{Convergence of Fourier Series}
\label{app-zen}

\noindent Let $\mathrm{g}(x)$ be a function that is continuous in value and first derivative and piecewise-continuous and bounded in second and third derivatives, which we differentiate twice to give a function, $\mathrm{f}(x)$ that is piecewise-continuous and bounded in value and first derivative.

\begin{thm}[Harper]
Let $\mathrm{f} : \RR \to \CC$ be periodic of period $T>0$, and let $\{ t_j \}$ be a finite dissection of $[0,T]$:
\[
0 = t_0 < t_1 < t_2 < \ldots < t_N = T.
\]
Suppose that $\mathrm{f}$, $\mathrm{f}'$ are continuous at all points of $[0,T] \setminus \{ t_j \}$ (where $\mathrm{f}'$ denotes the differential w.r.t $x$), and that
\begin{equation} \label{eq:f_dashed_integrable}
\int_0^T |\mathrm{f}'(x)| dx = \sum_{j=0}^{N-1} \int_{t_j}^{t_{j+1}} |\mathrm{f}'(x)| dx  < \infty.
\end{equation}
(Note that this makes sense even though $\mathrm{f}'$ need not exist at $t_j$. Also note that this easily implies that $\lim_{x \tendsto t_j^+} \mathrm{f}(x)$ and $\lim_{x \tendsto t_{j+1}^-} \mathrm{f}(x)$ exist).

Then the Fourier coefficients of $\mathrm{f}$ for the interval $[0,T]$ decay like $1/|n|$.
\end{thm}
\begin{proof}
For convenience we use the exponential form of the Fourier series. Let $d_n$ be the coefficient of the $n^{th}$ component of the exponential Fourier series, which is defined:
\begin{equation}
\label{sjh_eq_appb}
d_n = \frac{1}{T}  \int_{0}^{T} \mathrm{f}(x) \exp(-i  n \omega x) \mathrm{d} x
\end{equation}
Staring with the integral between $t_j$ and $t_{j+1}$, which we can integrate by parts:
\begin{equation}
\int_{t_j}^{t_{j+1}} \mathrm{f}(x) \exp(-i  n \omega x) \mathrm{d} x = -\frac{1}{i n \omega} \big[ \mathrm{f}(x) \exp(-i n \omega x) \big]_{t_j}^{t_{j+1}} +
\frac{1}{i n \omega} \int_{t_j}^{t_{j+1}} \mathrm{f}'(x) \exp(-i n \omega x)  \mathrm{d} x,
\end{equation}
so that
\begin{equation}
\left| \int_{t_j}^{t_{j+1}} \mathrm{f}(x) \exp(-i n \omega x) \mathrm{d} x \right| \leq 
\frac{1}{|n \omega|} \left( |f(t_j^+)| + |\mathrm{f}(t_{j+1}^-)| + \int_{t_j}^{t_{j+1}} |\mathrm{f}'(x)| \mathrm{d} x \right).
\end{equation}
\textbf{Note:} we need the one--sided limits $\lim_{x \tendsto t_j^+} \mathrm{f}(x)$ and $\lim_{x \tendsto t_{j+1}^-} \mathrm{f}(x)$ to exist,
which they easily do because of~\eqref{eq:f_dashed_integrable}, as remarked above.

There are \emph{finitely many} such terms, so summing gives
\begin{equation}
\left| \frac{1}{T} \int_{0}^{T} \mathrm{f}(x) \exp(-i n \omega x) \mathrm{d} x \right| \leq \frac{1}{|n \omega| T} \left( A + \int_0^T |\mathrm{f}'(x)| \mathrm{d} x \right) = \frac{B}{|n \omega|}
\end{equation}
for some $A, B>0$ (which of course depend on $\mathrm{f}$). Referring back to (\ref{sjh_eq_appb}) we can see that we have shown that $|d_n|$ indeed decay like $1/|n|$.
\end{proof}

If we now integrate the Fourier series of $\mathrm{f}(x)$ twice w.r.t $x$ to get the Fourier series of our original function, $\mathrm{g}(x)$, then we can see that we get Fourier coefficients decaying as $1/|n|^3$. That is because the $n^{th}$ term in the Fourier series is $d_n \exp(i  n \omega x)$, and $\int d_n \exp(i  n \omega x) \mathrm{d} x = \frac{d_n}{i n \omega} \exp(i  n \omega x)$ (and likewise for the second integration).

\end{document}